# Fast Point-Feature Label Placement for Dynamic Visualizations

Kevin Mote
Pacific Northwest National Laboratory,
Washington State University

**ABSTRACT**

This paper describes a fast approach to automatic point label de-confliction on interactive maps. The general Map Labeling problem is NP-hard and has been the subject of much study for decades. Computerized maps have introduced interactive zooming and panning which has intensified the problem. Providing dynamic labels for such maps typically requires a time-consuming pre-processing phase. In the realm of visual analytics, however, the labeling of interactive maps is further complicated by the use of massive datasets laid out in arbitrary configurations, thus rendering reliance on a pre-processing phase untenable. This paper offers a method for labeling point-features on dynamic maps in real time without pre-processing. The algorithm presented is efficient, scalable, and exceptionally fast; it can label interactive charts and diagrams at speeds of multiple frames per second on maps with tens of thousands of nodes. To accomplish this, the algorithm employs a novel geometric de-confliction approach, the "trellis strategy," along with a unique label candidate cost analysis to determine the “least expensive” label configuration. The speed and scalability of this approach make it well-suited for visual analytic applications.

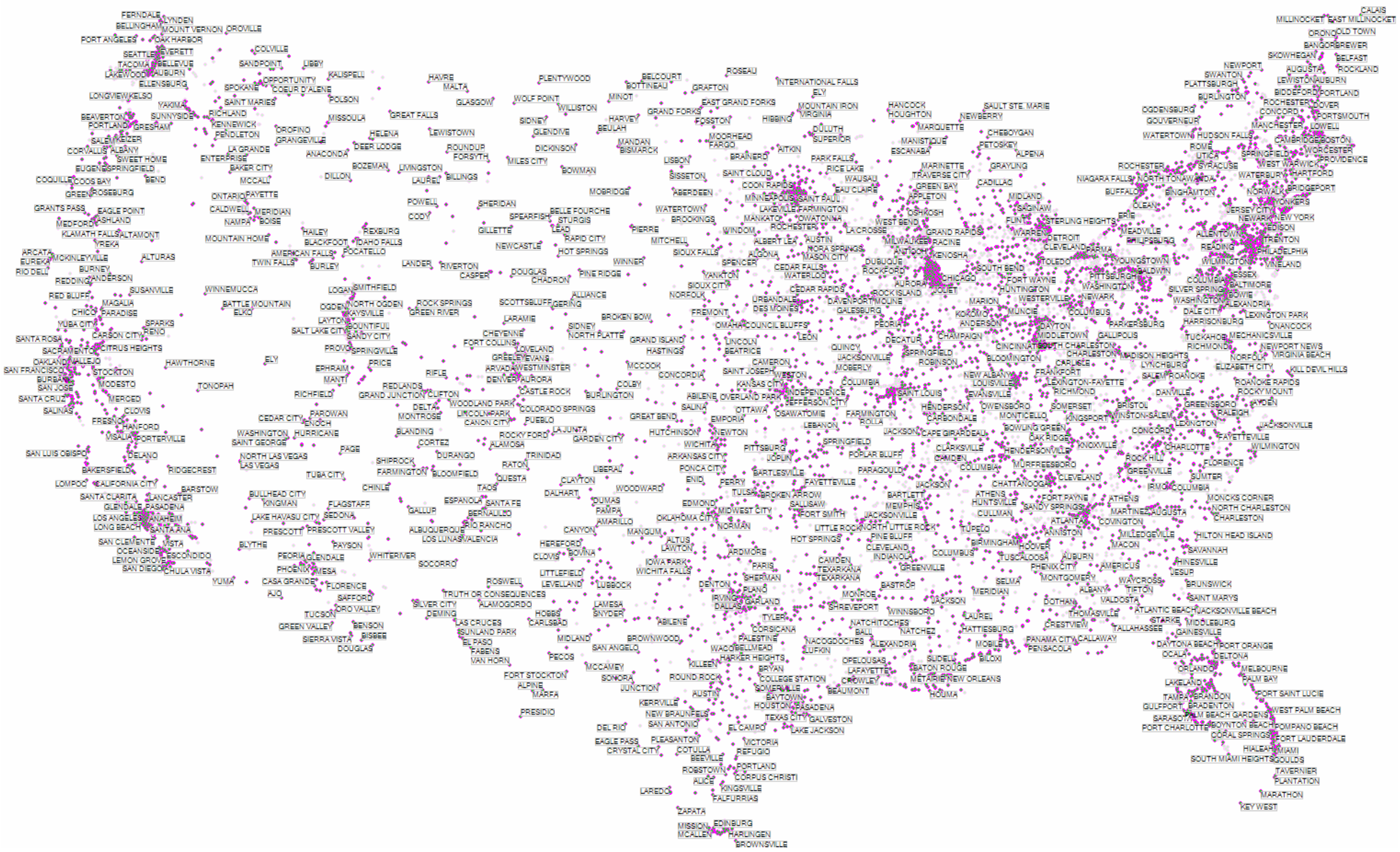

**Figure 1:** A view of over 10,000 cities in the U.S. displaying 800 labels that were computed in less than 0.07 seconds.

## 1 INTRODUCTION

The problem of placing non-conflicting labels on maps is simple in its specification: attach a label unambiguously to every point or feature on a map without allowing the labels to overlap each other or other features. Although it is theoretically feasible to solve the problem by simply enumerating all possibilities, the combinatorial explosion of this approach renders it intractable. Many different variants of the label placement problem have in fact been shown to be NP-hard [2, 11, 21], even in the one-dimensional case [26]. Research is therefore directed at finding the best approximation of an optimal solution. A decade ago the problem was identified as one of the most important areas of research in Discrete Computational Geometry [4]. It has long been a critical issue for such endeavors as graph and map drawing, as well as geographic and avionic information systems.

With the advent of information visualization and visual analytics, the challenge has only intensified. Visual analytic software brings unique demands: agile interactivity (including zooming and panning in 2D, or even rotation in 3D), unpredictable configurations (produced on the fly), and immense feature sets (multiple thousands of nodes). By way of example, the Starlight Visual Information System can generate a variety of models, or "views", consisting of multiple feature points displayed graphically in clusters, network graphs, or geospatial distributions, etc.[25] These views, generated by user-specified queries in a highly interactive environment, can easily contain tens of thousands of points or more. Besides Starlight, many other systems exist which spatially represent the text documents of large databases (e.g.,[28, 30]). In such environments the need to preserve contextual awareness through rapidly produced and readable labels is critical. But despite many recent advancements, most computational solutions for automatic label placement have proved inadequate for the exacting demands of interactive and dynamic visualizations.

The general problem of Map Labeling can be sub-divided into various categories, depending on the type of feature being labeled, whether points [5, 27], lines [18, 34], or polygonal areas [8, 19]. This paper will address the point label problem for dense point-clouds (cf. Figure 4). The remainder of the paper is organized as follows: Section 2 provides a brief overview of the literature. Section 3 specifies the precise problem. Section 4 describes an algorithmic solution. Section 5 outlines the results obtained by applying the algorithm to maps with dense feature sets. Finally, Section 6 summarizes the findings and offers suggestions for further study.

## 2 RELATED WORK

Research on the problem of automatic label placement stretches back for decades and has proceeded along a number of parallel tracks; researchers in cartography, computational geometry, and geographic information systems have been the principal developers. Accordingly, a wide variety of strategies have been applied over the years: from greedy and exhaustive rules-based approaches, to "divide and conquer," gradient descent, simulated annealing, genetic algorithms, linear/integer programming techniques, tabu search, ant colonies, and many more. In addition, the research has addressed various different label types in regard to their size, shape, and configuration. For example, labels have been modeled as squares, circles, fixed-height rectangles, or elastic frames. Other parameters to the problem include label orientation (axis-aligned or arbitrary), the number of label candidates per feature (2, 4, 8, or more), candidate space configuration (fixed, slider, or non-adjacent with leader-lines), optimization goal (size- or count-maximization), metric of success (speed or thoroughness), user-dependence (automated or guided), and application-domain (cartography, geographic information systems, information visualization systems). The influence of cartographic principles on information visualization is discussed in [28]. For comprehensive information on map labeling, see A. Wolff's excellent website and associated bibliography [33].

Historically, most of the literature has centered on the cartographic problem of finding the *best* solution: producing a map with the most labels, for example. Little consideration was given to the *speed* of the algorithm. Within the past few years, however, new breakthroughs have been made to provide faster algorithms for label generation. This has become increasingly important in the age of dynamic, computer-generated maps and displays. For example, until a few years ago, the best solutions in the scientific literature operated on the order of seconds or minutes, which is acceptable when generating a static map for printing. It is too slow, however, for dynamic maps that allow zooming and panning. In these applications, the label positions must be re-calculated with every change of scale or scope. An acceptable algorithm, therefore, must operate in real-time (i.e., "on the wheel of a mouse") for such an application to be useful.

Wagner et al. were among the first to provide a faster approach, suggesting heuristic rules that brought a significant improvement over previous approaches but that were still too slow for large sets, measuring in minutes for large sets [32]. Petzold followed by providing real-time labeling, albeit with a high pre-processing cost [23]. A so-called "Real-Time Method," intended to be suitable for personal navigation applications, was described in [37], but the authors concede that "for real-time applications . . . we find our current implementation not efficient enough." In a related paper an algorithm is presented that is indeed "fast enough for most real-time map applications," however the authors offer this efficiency only for datasets with about 100 points[15]. A recent approach utilizing a "greedy randomized adaptive procedure" provides good results but requires well over a minute for datasets of 1000 points and does not test sets larger than this [7]. Most promising, perhaps, is a graph-theoretic algorithm described in [27] that offers a theoretical runtime of O( $n\sqrt{n}$ ). The authors provide tabulated results indicating a sub-second speed for instances of size 1000. Much larger sets require multi-second time, however, and the approach does not appear to lend itself to feature prioritizing or label placement preference. A number of other recent approaches, all with a common strategy of using an expensive pre-processing phase, will be discussed further in Section 4.1, below.

Dr. Herbert Freeman, whose work in this field spans decades, offers an excellent synopsis of the last 25 years of progress on automated cartographic text placement systems. He concludes his overview with the promise that “the day of one-second quality labeling of an electronically displayed…map or…chart is not far off” [12].

## 3 Statement of Problem

The goal of this research is to identify a method for labeling the features on dynamic maps in real time without a pre-processing stage. The term “maps” will be used in this paper to refer to any chart, view, or diagram produced by information visualization and visual analytic software, in which the main features (i.e., elements to be labeled) are represented as *points*. These “point-features” will be referred to henceforth as simply “features”. The set of all features will be called the map set. These maps are considered “ad hoc” in that they are generated on the fly in unpredictable configurations and require a labeling method that can be calculated instantly. Moreover, the data sets can be massive: ranging up to tens of thousands of features or more.

The precise scope of the problem can be defined as follows: Given a map with a pre-defined set of prioritized features, swiftly find the largest possible set of non-intersecting, axis-aligned rectangular labels, giving preference to the highest-prioritized features and to the cartographically preferred candidate locations. Each label must: (a) touch its referent (parent) feature at one of its four corners, (b) obscure as few other features as possible, and (c) obscure no other labels.

In seeking to address this problem, the following design decisions have been made in the algorithm to address the specific challenges posed by dense feature-sets in interactive displays:

***It does not restrict labels from obscuring other features in the view.*** This approach represents a significant departure from most previous schemes, but it is clearly unavoidable when working with the dense clusters produced by visualizations of massive data sets. The on-screen density of these visualizations often provides little or no white space for non-conflicted labels. To prohibit occlusion of features in these cases would mean that very few features (if any) would be labeled at all. Worse, because the highest-prioritized features are often buried deep in clusters, only the marginal outliers would be labeled, in direct opposition to our stated goal of priority preference. That being said, this algorithm attempts to allow the fewest features possible to be obscured, and then, only by higher priority feature labels.

***It is optimized for uniformly-sized labels.*** The trellis strategy, as described in Section 4.1 below, is designed primarily for labels of equal size. Nevertheless, non-uniform labels can be accommodated, albeit with diminishing quality. For example, users could specify that longer-than-average labels may be truncated, or alternatively, may be allowed to overlap other labels by a certain percentage. Experimentation has demonstrated that a combination of these two options can produce very acceptable and readable labelings with variably sized labels.

***It exploits priority-ranked features.*** It is common in visual analytic applications that the individual features are associated with a specific value representing their “importance” to the view. This value may be derived or assigned. For example, if the view represents a geographic map with features pinpointing the cities, the features might be ranked according to the population of those cities. Or if the view was a complex scatter-plot displaying the results from a web search, the feature priority might correspond to their weight, Google™ page-rank, chronological order, or any other ordered numeric attribute of the data. The use of such preference information, if available, can greatly increase the efficiency of the algorithm. See [2] for similar use of priority rankings.

***It prefers speed over optimal label configurations.*** This design decision is based squarely on the interactive nature of dynamic maps. The assumption here is that any sub-optimal configuration or indistinct labeling can in most cases be disambiguated with a minor amount of user interaction (e.g., zooming). While this allows us to relax our expectations slightly with regard to perfect labelings, experimental results demonstrate that quality is only marginally degraded.

***It is based on a label model of size four***. A "label model" refers to *the number of possible positions,* or "*candidates,*" in which a label may be located around a given feature. Many mapping utilities and geographic information systems use label models offering more than four candidates per feature. Some offer the "slider" model, where labels are allowed to "slide" around the feature so that the feature may appear anywhere on the boundary of the label (cf. [31, 38]). However, the contention here is that such extended label models are inappropriate for dense visualizations. Given the magnitude of screen-space density in high-volume visualizations, particularly when labels are allowed to over-post features as discussed above, referent-ambiguity increases uncontrollably as the label model grows. Because so many features may feasibly appear on or near the boundary of a label, only by constraining the labels so that their referent feature appears at one of their corners can we preserve appropriate contextual awareness. Similarly, the "excentric" model, where labels are attached by leader lines to their associated features [3, 10], is not particularly suited for high-density displays.

***It recognizes cartographic preference***. The relative preference of each of a label's four candidates is determined by the aesthetic criteria established by cartographers. Described in [17], these informal rules establish the relative value of each candidate. Typically, the upper-right candidate is preferred, followed by the lower-right, upper-left, and lower-left, in that order. Other aesthetic issues are discussed and evaluated in [16].

## 4 ALGORITHM: LABEL SELECTION BY CONFLICT- EXPENSE ESTIMATION

This section presents a solution to the stated problem, introducing a rapid, yet effective method of determining and selecting what will be referred to as the "least expensive" label configuration. The algorithm is divided into the following three conceptually distinct steps:

- Conflict detection
- Expense calculation
- Label candidate selection

In implementation, these steps would actually be merged. *First*, analyze each feature in the set, testing it against all features that lie in close proximity, and create what is known as a “conflict graph” to keep record of label candidate intersections, or “conflict partners”. Although this step would normally be quite expensive, an efficient approach—the trellis strategy—will be introduced to expedite these calculations. *Second*, pass through the set of label candidates and, for each one, calculate an associated *expense*. This expense will be described in detail below but is essentially a function of the sum of the label’s conflict partners along with their associated priorities, among a few other factors. *Finally*, pass through the set of features, in descending priority order, and for each feature, select its “least-expensive” label candidate, while de-selecting that feature’s remaining candidates. In essence, this algorithm uses an informed but greedy approach to approximately minimize the expense of the total label set.

By computing each candidate in priority order, we can have high confidence that a large majority of the highest priority features will be labeled at the default zoom level, assuming an approximately uniform distribution of the visualization data features. This system does not offer demonstrable asymptotic guarantees of performance, as it claims only to be a useful approximation. Nevertheless, the author’s implementation provides strong evidence that this approach is fast, effective, and reliable in many real-world scenarios. Furthermore, it should be reiterated that this system requires no extended pre-processing stage, making it ideal for information visualization and visual analytic applications in which maps, diagrams, and charts are constructed on the fly. This is particularly true of three-dimensional feature maps, for which preprocessing every conceivable orientation is entirely unfeasible.

### 4.1 Step One: Conflict Detection

One of the most common approaches in the literature to solving the point-label problem uses a concept referred to as a “conflict graph [29]”, an “overlap graph” [19], or a “label graph” [27]. These terms refer to a graph whose nodes correspond to labels and whose edges correspond to intersections between labels. Using this model, the objective of the label placement problem is to find the maximum independent set of the conflict graph [1]. The concept of a conflict graph can also be understood more generally, in a non-graph-theoretic sense, as an “adjacency matrix” storing intersection information for all feature pairs. This paper generalizes the concept of a conflict graph, defining it simply as a list of conflict partners associated with each feature. This list will act as a look-up table to determine the total number of conflicting pairs of label candidates or, more specifically, the total number of occluded labels for any given candidate, and to keep record of the label candidates that must be removed due to occlusion.

Although constructing the conflict graph is typically quite expensive, this step is often disregarded in the literature. The conflict graph is quite often assumed as input to many algorithms (see for example [2, 32, 35]). Such papers generally describe how to determine or identify the largest independent set of label candidates, given a *pre*-determined conflict graph. This approach is appropriate if the size of the dataset is relatively small, or if the pre-processing time available for conflict graph construction is irrelevant. Unfortunately, neither of these assumptions holds true in the realm of dynamic information visualization applications.

Indeed, even though many of these approaches are offered for their efficiency, this overlooked step can often be the bottleneck of performance.

In fact, some more recent papers have sought greater efficiency in label-selection by *intensifying* this requirement, rather than abbreviating it. In [22, 23], for example, an approach is described for determining a "reactive conflict graph." This graph stores information about all potential conflicts at all zoom levels. It is constructed during a several-minute-long "pre-processing" phase and rapidly accessed during an interactive phase. In [24], a line-stabbing approach to conflict graph generation is used, building low-height hierarchies to better ensure the quality of the final layout at multiple zoom levels. This approach has the advantage of offering asymptotic guarantees of performance but requires $O(n^2)$ time for two-dimensional maps. Likewise, interactive speed is achieved in [2], but "all of the selection and placement decisions are moved into the preprocessing stage." These approaches, with their reliance upon a time-consuming pre-processing stage, make them unsuitable options for the ad hoc map generation of dense information visualization applications.

Conceivably, one could attempt to eliminate conflict graph construction altogether by simply estimating the number of conflicts for any given feature. By using, for example, a modification of the "trellis" approach described in this paper such estimates could indeed turn out to be reliable and sufficient. However, such an approach does not address the second purpose of the conflict graph mentioned above; that is, it provides no way to determine which specific labels must be removed due to occlusion. For this reason, a conflict graph of some form must be generated and must be available and accurate at all zoom-levels and orientations of the feature set. Although the construction of a *globally* complete graph is not feasible for any fast de-confliction algorithm, the ideal solution must at least include a strongly reliable *approximation* of the graph. This necessity will require a considerably more sophisticated and informed approach to conflict graph generation than the naïve $O(n^2)$ method of testing all candidates against all others. The following section will describe in detail a strategy that can in principle offer an efficiency gain of nearly three orders of magnitude over the elementary approach. The dual goal of this new approach is to accelerate the detection of intersecting label candidates, and to avoid all unnecessary testing of feature pairs.

#### 4.1.1 The Trellis Strategy

The strategy employed here is based on a geometric subdivision of the feature set in screen-space. This subdivision, called a "trellis," can be conceived as a two-dimensional array of equally sized cells, sub-dividing the screen-space view into rows and columns. Each cell in the trellis has an associated list of all the features that are located in that cell's defined pixel space. The trellis, of course, would never actually be rendered in any visualization, but if it were it would resemble a cross-hatch lattice of horizontal and vertical lines—like a garden trellis. This trellis is similar in many respects to the "grid of buckets" described in [2]. The approach described here, however, exploits the characteristics of a slightly more sophisticated subdivision of map space.

One very significant and strategically determined characteristic of the trellis is the size and aspect ratio of the cells. Each cell is defined as a "*quarter-region*"; that is, it is exactly one-fourth the area associated with a label, the size of which we assume to be fixed and constant (cf. Section 3). It also shares the same aspect ratio of the label. In other words, if one were to slice a label in half both vertically and horizontally, the resulting four quarter regions define the trellis cell (cf **Figure 2**). (Henceforth, "trellis cell" and "quarter-region" will be used interchangeably.) The purpose for this particular size and shape will be explained shortly. All the cells in the trellis are, hence, identical, with the possible exception of the rightmost column and the bottom-most row of cells, which may be smaller if view-space constrains them. For a view of size 1500x1000 pixels, with labels of size 150x20, there would be 2000 cells in the trellis.

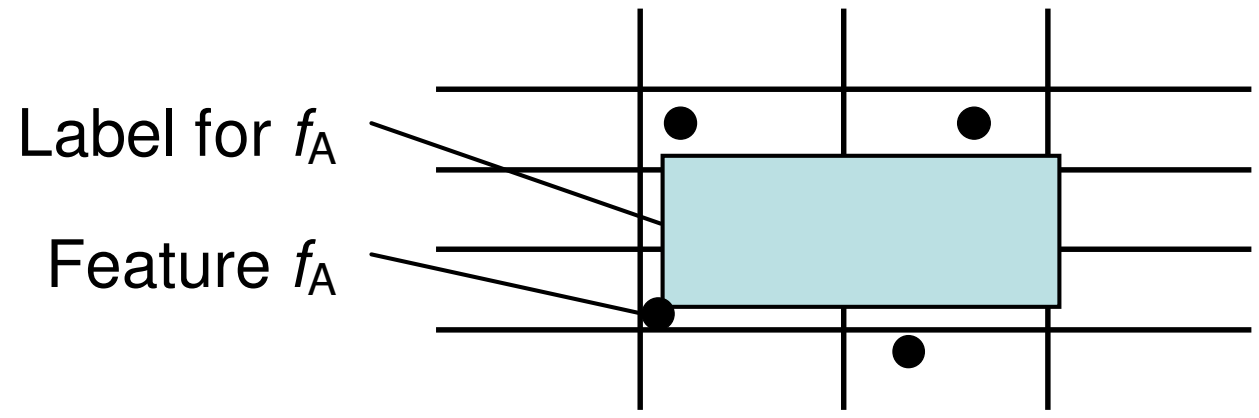

**Figure 2:** A portion of the trellis. The label of feature point $f_A$ is displayed to demonstrate that each trellis cell is one-fourth the size of the labels. Other features are shown in the neighborhood of $f_A$.

The trellis is populated by testing each feature $f$, performing the following two calculations to determine which row and column of the trellis contains the feature.

$$\text{Trellis}_{\text{column}}(f) = \text{floor}(f_x/\text{quarter_region_width})$$
$$\text{Trellis}_{\text{row}}(f) = \text{floor}(f_y/\text{quarter_region_height})$$

where $f_x$ and $f_y$ represent the x and y pixel coordinates of $f$, respectively, relative to the view window. Hereafter the trellis cell in which feature $f_A$ lies will be referred to as CELL($f_A$). The expense of this series of calculations is, of course, linear in $n$.

Once we have iterated the feature set, every trellis cell is associated with a list of the features that appear within its boundaries. Conversely, every feature has an associated trellis coordinate. By taking advantage of the specific geometric structure and layout of the trellis, the conflict graph can be generated in an extremely efficient way.

#### 4.1.2 The Trellis Advantage

The first and most obvious advantage offered by the trellis is the ability to specifically and instantly define a neighborhood around any given feature. As a result, we will not have to test each feature against every other feature in the set (an $n^2$ operation). Instead, we can limit our conflict search so that, for any given feature, we will test it only against those features that are located in the immediately neighboring cells. Specifically, we will inspect only those cells that have a possibility of containing conflicting label candidates. This approach was similarly described in [2].

By the nature of the design of the trellis dimensions, it is a simple task to determine which cells have potential conflicts. Because quarter-regions are ¼ the size of a label, it may be observed that a label candidate of feature $f_A$ can conflict with a label candidate of feature $f_B$ if and only if $f_B$ resides in a four-cell radius of $f_A$, or no more than four rows or four columns away from $f_A$.

Such a radius defines a 9x9 array of cells, centering on the home, or "parent" cell of $f_A$ (cf. **Figure 3**). This array will hereafter be referred to as the "neighborhood" of $f_A$, and features within those cells, as "neighboring features" of $f_A$. All features outside of this neighborhood can be disregarded with respect to $f_A$; their label candidates have no possibility of conflicting with those of $f_A$ beyond a coincident edge. On the other hand, the label candidates of all neighboring features of $f_A$ are regarded as potential conflict partners with the candidates of $f_A$.

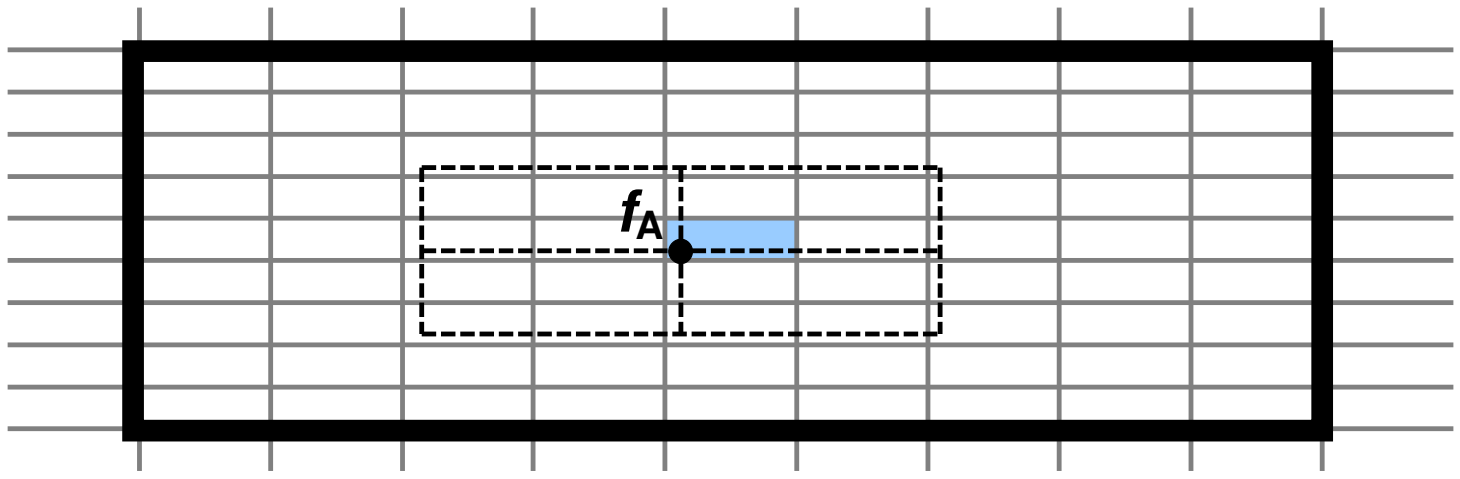

**Figure 3:** A portion of the trellis displaying the entire neighborhood (bold line) of feature $f_A$. Four candidate label locations are shown (dashed lines). Note that regardless of where $f_A$ is located in its parent cell (blue), it is guaranteed not to conflict with the labels of any features outside this neighborhood.

Referring to the example view mentioned above, we can see that this has reduced the target range of test cells from 2000, which was the number of cells in the entire view, to 81. If at least a nominally uniform distribution of features is assumed, it can be seen that this strategy has reduced the number of potential conflict tests by roughly one order of magnitude.

The benefit of subdividing our space into quarter-regions is that it can significantly reduce the number of conflict tests necessary. For instance, consider a feature $f_A$ in CELL($f_A$), and a second feature $f_B$ in one of the neighboring cells. Because CELL($f_B$) is in the neighborhood of CELL($f_A$), their label candidates are potentially in conflict. Imagine next that CELL($f_B$) is exactly four cells to the left of CELL($f_A$). Due to this fixed Cartesian relationship we can narrow the list of potential conflicts, thus eliminating the need to test all of them. The complete list of possible conflicts between the label candidates of $f_A$ and $f_B$ in this case are as follows:

[A0:B2, A0:B3, A1:B2, A1:B3]

We therefore only have to test those particular candidates for conflicts rather than all 16 possible conflict pairs. Tables 1-3 delineate precisely all the possible conflicts that can occur in each cell in the neighborhood of a central feature. Moreover, as we continue to exploit the known Cartesian relationships between the cells, we gain an additional advantage. Note that, typically, conflict graph generation requires the detection of rectangle intersections, which is not necessarily an inexpensive test. Many approaches to the general problem of rectangle intersection-pair identification have been published within the field of computational geometry (e.g., [6, 14]), but the problem itself is a difficult one. Even the simple case of testing a single pair of rectangles is fairly incompressible, with essentially no way to abridge it beyond four atomic operations, such as

IntersectionExists(rectA, rectB) =

```
((rectA.left < rectB.right) && (rectA.right > rectB.left) &&  (rectA.top > rectB.bottom) && (rectA.bottom < rectB.top))
```

Fortunately, however, the use of the trellis method eliminates the need to perform any rectangle intersection tests at all. Consider the example described above, in which $f_B$ lies 4 cells to the left of $f_A$, and four possible candidate intersections exist between them. Rather than testing each one of the four pairs for intersection, we will exploit the information provided to us by the geometric orientation within the trellis. In this particular case, then, we merely have to test whether the distance from $f_A$ to $f_B$ is greater than the width of two labels, in which case no conflicts are possible. If the distance is less than two label-widths then we additionally test whether $f_A$ is higher (in the *y*-direction) than $f_B$. The result of these two tests determines precisely which of the four possible label conflicts actually occur.

We can use similar observations in each cell of a given neighborhood to greatly reduce the number of required tests. Table 1 represents one quadrant of the trellis neighborhood and outlines all the tests required for each cell in that area. As the table demonstrates, none of the cells in the entire 81-cell neighborhood require more than two atomic tests; many require only one; and a few require none at all. If one feature were to reside in every cell of the neighborhood of $f_A$ there would then be at most 90 tests required to build the complete conflict graph with respect to $f_A$. Note that the simplest rectangle intersection approach would require testing all four of $f_A$'s label candidates against the four candidates of each of its 80 neighbors, or 4x4x80=1280 total tests. If the cost of the four-operation intersection test were included, the total count would approach 5000 atomic operations. Contrast this with the 90 operations of the trellis strategy, and it can be seen that the efficiency has been increased by a factor of nearly 500:1. The trellis strategy has, hence, reduced the cost of the problem by a second order of magnitude, for the uniformly-distributed case.

This demonstrates the tremendous benefit offered by the trellis method. For the low (O(*n*)) cost of a single pass through *F* to populate the trellis, a potential speed-up of nearly 2½ orders of magnitude has been achieved over a rudimentary approach to conflict graph generation. In addition, another characteristic of the trellis will be used shortly in calculating the values of each label candidate.

**Table 1: The Trellis Neighborhood Test Outline** (showing only the upper-left quadrant of the neighborhood, due to space constraints). The central cell is bottom right (in blue). The other three quadrants (not shown), are symmetric, but not identical, to this one. Each cell is defined as a "quarter-region" being one fourth the size of an actual label. The specific tests, in bold, are explained more fully in Table 2. The result of each test is a particular label conflict configuration, abbreviated with codes such as α1 or β1, and delineated precisely in Table 3.

| | | | | |
|---|---|---|---|---|
| **1** -4/-4<br>**YΔ>2L?**<br>/ \ <br>α0 : **XΔ>2L?**<br>/ \ <br>α1 : β1 | **2** -3/-4<br>**YΔ>2L?**<br>/ \ <br>α1 : β1 | **3** -2/-4<br>**YΔ>2L?**<br>/ \ <br>α1 : **XΔ>1?**<br>/ \ <br>β1 : γ13 | **4** -1/-4<br>**YΔ>2L?**<br>/ \ <br>α1 : γ13 | **5** 0/-4<br>**YΔ>2L?**<br>/ \ <br>α1 : **$X_A>X_B$?**<br>/ \ <br>γ13 : γ31 |
| **10** -4/-3<br>**XΔ>2L?**<br>/ \ <br>α0 : β1 | **11** -3/-3<br>**No test** | **12** -2/-3<br>**XΔ>1L?**<br>/ \ <br>β1 : γ13 | **13** -1/-3<br>**No test** | **14** 0/-3<br>**$X_A>X_B$?**<br>/ \ <br>γ13 : γ31 |
| **19** -4/-2<br>**XΔ>2L?**<br>/ \ <br>α0 : **YΔ>1L?**<br>/ \ <br>β1 : γ10 | **20** -3/-2<br>**YΔ>1L?**<br>/ \ <br>β1 : γ10 | **21** -2/-2<br>**XΔ>1L?**<br>/ \ <br>**YΔ>1L?** : **YΔ>1L?**<br>/ \ / \ <br>β1 : γ10 γ13 : δ1 | **22** -1/-2<br>**YΔ>1L?**<br>/ \ <br>γ13 : δ1 | **23** 0/-2<br>**YΔ>1L?**<br>/ \ <br>**$X_A>X_B$?** : **$X_A>X_B$?**<br>/ \ / \ <br>γ13 : γ31 δ1 : δ3 |
| **28** -4/-1<br>**XΔ>2L?**<br>/ \ <br>α0 : γ10 | **29** -3/-1<br>**No test** | **30** -2/-1<br>**XΔ>1L?**<br>/ \ <br>γ10 : δ1 | **31** -1/-1<br>**No test** | **32** 0/-1<br>**$X_A>X_B$?**<br>/ \ <br>δ1 : δ3 |
| **37** -4/0<br>**XΔ>2L?**<br>/ \ <br>α0 : **$Y_A>Y_B$?**<br>/ \ <br>γ01 : γ10 | **38** -3/0<br>**$Y_A>Y_B$?**<br>/ \ <br>γ01 : γ10 | **39** -2/0<br>**YΔ>1L?**<br>/ \ <br>**$Y_A>Y_B$?** : **$Y_A>Y_B$?**<br>/ \ / \ <br>γ01 : γ10 δ2 : δ3 | **40** -1/0<br>**$Y_A>Y_B$?**<br>/ \ <br>δ0 : δ1 | **41** 0/0<br>**$X_A>X_B$?**<br>/ \ <br>**$Y_A>Y_B$?** : **$Y_A>Y_B$?**<br>/ \ / \ <br>δ0 : δ1 δ2 : δ3 |

Trellis "neighborhood" coordinate

Conditional tests;
Yes: branches left
No: branches right

Conflict configuration. (See Table 3 for delineation of pair conflicts)

**Table 2:** Legend defining the tests used in Table 1

**Conditional Tests:**

**YΔ>1L?**
→ abs($f_A$.y_coord - $f_B$.y_coord) > Label_height

**YΔ>2L?**
→ abs($f_A$.y_coord - $f_B$.y_coord) > 2*Label_height

**$Y_A>Y_B$?**
→ $f_A$.y_coord > $f_B$.y_coord

**XΔ>1L?**
→ abs($f_A$.x_coord - $f_B$.x_coord) > Label_width

**XΔ>2L?**
→ abs($f_A$.x_coord - $f_B$.x_coord) >2*Label_width

**$X_A>X_B$?**
→ $f_A$.x_coord > $f_B$.x_coord

| Types: | Possible Label Pair Configurations | | | |
|---|---|---|---|---|
| α | 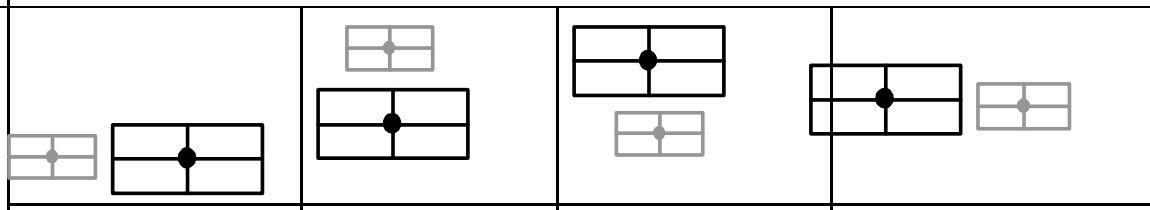 | | | |
| | α0 | α1 | α2 | α3 |
| CPs: | Ø | Ø | Ø | Ø |
| β | 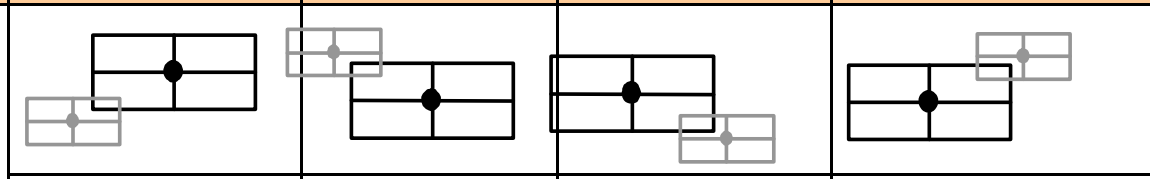 | | | |
| | β0 | β1 | β2 | β3 |
| CPs: | A0:B3 | A1:B2 | A2:B1 | A3:B0 |
| γ | 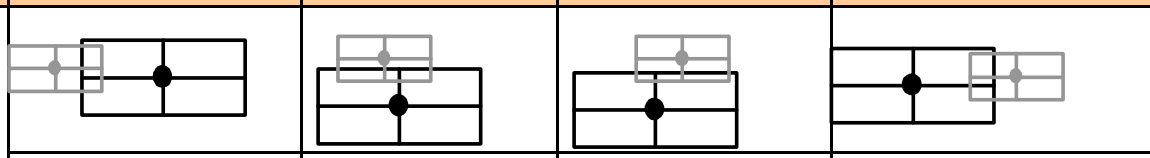 | | | |
| | γ10 | γ13 | γ31 | γ32 |
| | A0:B2, A1:B2, A1:B3 | A1:B0, A1:B2, A3:B2 | A1:B0, A3:B0, A3:B2 | A2:B0, A3:B0, A3:B1 |
| | 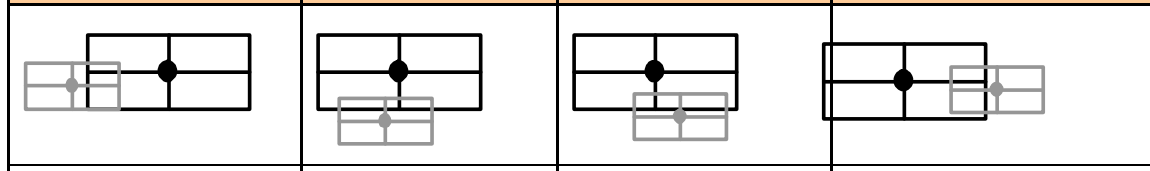 | | | |
| | γ01 | γ02 | γ20 | γ23 |
| CPs: | A0:B2 A0:B3 A1:B3 | A0:B1 A0:B3 A2:B3 | A0:B1, A2:B1, A2:B3 | A2:B0, A2:B1, A3:B1 |
| δ | 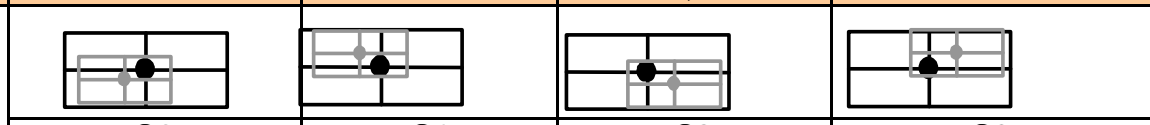 | | | |
| | δ0 | δ1 | δ2 | δ3 |
| CPs: A0:B0, A1:B1, A2:B2, A3:B3 | A0:B1, A1:B3, A0:B2, A2:B3, A0:B3 | A1:B3, A3:B2, A1:B2, A0:B2, A1:B0 | A2:B0, A3:B1, A2:B1, A0:B1, A2:B3 | A3:B0, A1:B0, A3:B1, A2:B0, A3:B2 |

**Table 3:** This table represents every possible configuration between a given pair of features, along with the corresponding "conflict pairs" (CPs) among two label candidates, A and B. (Feature $f_B$'s candidates are smaller for illustration only). The notation [A2:B0] denotes: "$f_A$'s candidate #2 conflicts with $f_B$'s candidate #0." The configurations are grouped in four types (α, β, γ, δ). In α–configurations no conflict occurs between label candidates, whereas in δ–configurations all four of $f_B$'s candidates are obscured.

### 4.2 Step Two: Cost Analysis

Having established a conflict graph for each feature, the objective of the second step of the algorithm is to determine the least expensive label candidate position for each feature, among its four options. Each label candidate has an inherent value based on the priority of its referent feature. (Recall that the features are prioritized or prioritized in order of preference.) The goal of the algorithm is simply to maximize the sum of all the values in a set of non-occluded candidates.

By way of illustration, imagine a deck of playing cards randomly scattered face-up on a small table. The face value of the card represents its priority and the suits represent the four label candidates of a given feature. Cards resting on top of other cards represent conflicting label positions. Assume that the cards were dealt in sorted order, so the kings are on top. The task is to choose at most one suit for each card value, while removing all cards that are partially obscured by higher valued cards. Once a suit is selected for a given card value all cards it "conflicts" with (the ones directly under it) must be removed from the table. The goal is to produce a set of non-occluded cards with the highest possible total value. By using this analogy it is fairly easy to see that one effective strategy would be to begin with the four kings and select the one which rests on the "least expensive" pile of cards. We call this selected card the least expensive candidate. Once this selection is made the other three kings are removed, and also all the cards that were under our selected king. We can then proceed in the same way with the remaining queens and so on. Determining which stack of conflicting cards is least expensive is simply a matter of finding the sum of the face value of those cards. This card game illustrates the strategy used by the algorithm of this paper. The goal, simply put, is to approximately maximize the set of all the values of non-occluded feature labels.

It should be noted that, although we have spoken of maximizing the expense of the map-set, this approach is not strictly an optimization algorithm. It is not claimed that the global maximum of the feature set expense will be found. The algorithm is essentially a "greedy" one and is indeed vulnerable to local maxima. Nevertheless, as will be demonstrated, this approach will provide a useful and reliable approximation of the optimal outcome. Our confidence in this approximation stems from the fact that we are processing our features in priority order. As we pass through our feature set in descending order, choosing the least expensive candidate for each feature, our guarantee is that no feature will remain unlabeled unless all of its candidates are occluded by *higher* priority labels. This approach is, of course, dependent upon an accurate assessment of the inherent value of each label. This value will now be examined more precisely, looking both at the static base value of a given label and, subsequently, at some dynamic run-time modifications to that value.

#### 4.2.1 Label Candidate Base value

As has already been noted, the default value of a given label is dependent directly on the priority of that label's referent feature. High priority features will obviously have more valuable labels. The model is further extended to include the "cartographic criteria" (described in Section 3) by assigning a higher value to the aesthetically preferred label candidates.

One drawback of this simplistic cost model is that it allows the possibility that a small cluster of low priority labels might "outweigh" a few higher priority labels. For example, consider a feature $f_A$ with two label candidates, $f_A^0$ and $f_A^1$, where $f_A^0$ conflicts with a high priority feature $f_B$ and $f_A^1$ conflicts with a group of low priority features. Ideally, $f_A^1$ should be selected because it obscures only low priority features. However, because the algorithm sums the value of the conflicts, $f_A^1$ may in fact be deemed more expensive, and consequently the high priority conflict partner $f_B$ would be occluded. To mitigate against circumstances such as this, the cost model requires the following adjustment. Taking $\text{VALUE}(f_0) = \text{BASE_VALUE}(f_0)$, iteratively update the value of all remaining features like so:

$$\text{VALUE}(f_i) = \text{VALUE}(f_{i-1}) + \frac{\text{BASE_VALUE}(f_i)}{n}, i = \{1, \ldots, n\}$$

Thus, the increment between any two consecutively prioritized features is changed from 1 to $i/n$ (where $\text{BASE_VALUE}(f_0) \approx i$). This effectively spreads the data to give increasingly greater weight to the higher priority features requiring, for example, twice as many mid priority label candidates to "outweigh" a high priority candidate.

#### 4.2.2 Label Candidate Value modifications

Up to this point, the value of a given label candidate has been fixed— determined at the time of view-initialization as a function of its referent feature priority and its label-model (aesthetic) preference. This value has so far been independent of the conflict graph. In order to increase the effectiveness of the cost-analysis approach, however, two specific modifications will now be introduced to adjust these values dynamically, in response to circumstances detected or produced in the conflict graph construction phase.

First, the value of a candidate will be adjusted upwards each time one of its siblings is occluded. The notion behind this is that as a feature loses its candidate labels due to occlusion, the remaining candidates become increasingly valuable. If a feature has only one non-occluded candidate left, it should be considerably more expensive to de-select it. We therefore increase the value of each label by an amount equal to the value of the occluded siblings. Hence, the value of occluding a neighbor's final candidate is equal to occluding all four of that neighbor's candidates bcause both cases result in the de-selection of an entire feature, preventing that feature from receiving a label. Therefore, the value of an "only child" should be equal to the sum of all the original candidates. By extending this logic, the value of any label should be incremented by the value of each lost sibling.

The second dynamic modification to our cost analysis addresses the interactive nature of the map. The notion here is that, given the user's ability to zoom, many label candidates that are occluded at one particular zoom level will be freed up as the user zooms in closer. Therefore, many of the features in the distant cells of the neighborhood of $f_A$ will no longer remain in the neighborhood at deeper zoom levels. Because of this, the value of a conflict partner of $f_A$ should increase with its *proximity* to $f_A$. In other words, features that are in the nearest cells to $f_A$ would potentially require more zoom steps to de-conflict, and should therefore be more expensive to occlude. The exact amount of the value adjustment in most cases should be a function of the magnification factor applied to each level of zoom. The author's own implementation uses the following adjustment schedule for the modified value:

$$\text{MODIFIED_VALUE}(f_B) = \text{BASE_VALUE}(f_B) * (\text{PROX_WT}*(5-\text{RAD_DIST}(f_A, f_B)))$$

where PROX_WT is the proximity weight factor (in this implementation it is .5), and RAD_DIST is the radial distance from $f_A$ to $f_B$ in cells (which ranges from 0 to 4). This distance is counted in rows or columns of separation, whichever is greater. This divides the neighborhood into four concentric rectangles around the parent cell of $f_A$ and increments the value between each progressively closer ring of cells by 50%. The ultimate effect of this adjustment is that the algorithm can "intuitively" determine the least expensive labels over several zoom levels. This will tend to mitigate inconsistency in label candidate choice from one zoom level to another.

### 4.3 Step Three: Label Selection

Once a weighted set of label candidates has been determined for each feature, the final label selection can be made by simply comparing the non-occluded candidates of each feature and selecting the least expensive one. All other candidates will be de-selected, which essentially disposes of them for the purposes of this algorithm, until the next viewer interaction reinitiates this entire process. Finally, using the conflict graph built in the first step, all conflict partners of the selected label are de-selected by occlusion. They are thereby removed from their respective referent feature's pool of label candidates.

Two additional refinements increase the algorithm's efficiency even further:

First, any feature $f_A$ that has already lost all of its candidates to occlusion from higher priority labels need not be tested against its neighbors. Doing so would be a waste of cycles, as none of $f_A$'s labels have any chance of being selected. In particularly dense sets, this can automatically prune a significant number of features.

Second, as a feature is being tested against its neighbors, it can safely ignore all higher priority features as it is guaranteed not to have any conflicts with them. If a conflict had existed in the initial configuration between a candidate of feature $f_B$ and that of a higher priority feature $f_{A,}$ it would have already been detected during $f_A$'s neighborhood traversal. In that case, either the conflicting candidate of $f_A$ was selected, and the occluded candidate of $f_B$ removed, or else a different candidate was selected for $f_A$ and the conflicting candidate was removed from contention. In both cases, the conflict is removed prior to the testing of $f_B$. Using this shortcut can cut the required number of tests in half.

As was stated previously, this algorithm was divided, for didactic purposes, into three steps: conflict graph generation, cost analysis, and label selection. In actual implementation these three tasks need not be performed separately. Rather, using the trellis strategy, the cost of each conflicting label candidate can be accumulated, the derived expense for the target candidates determined, and the optimal label candidate selected, all with one swift pass through the neighboring features.

## 5 Experimental results

The algorithm was applied to several data views generated by the Starlight Visualization System [ref]. These views had a dimension of 770x840. Tables 4 and 5 represents the time required to compute a full-map labeling for various (uniform) label sizes. The times recorded are the results of the author's implementation of the algorithm (in C++) on a Dell Xeon, 3.2Ghz machine running Windows, with 3GB of RAM. See Figure 4 for an example of a typically dense point cloud, along with its labels. The use of progressively smaller label sizes demonstrates the ability of the algorithm to calculate all map resolutions or zoom levels. Note that the choice of label size does not significantly increase the time required for computation. This was true regardless of the ratio of label-size to map. Table 5 displays the results for small labels, some smaller than a pixel. The purpose of this will be discussed further below, but it essentially demonstrates the ability of the system to process all zoom levels of an enormous map—without time penalty.

**Table 4**: Labeling speed (in seconds)

| # pts: | Label dimensions (w x h, in pixels) | | | |
|---|---|---|---|---|
| | 50x8 | 100x10 | 150x12 | 200x14 |
| 1K | < 0.001 | < 0.001 | < 0.001 | < 0.001 |
| 3K | 0.031 | 0.031 | 0.031 | 0.032 |
| 5K | 0.047 | 0.047 | 0.047 | 0.047 |
| 11K | 0.110 | 0.110 | 0.109 | 0.109 |
| 25K | 0.266 | 0.328 | 0.328 | 0.344 |
| 50K | 0.531 | 0.625 | 0.641 | 0.641 |
| 75K | 0.844 | 0.984 | 1.047 | 1.063 |

**Table 5:** Labeling speed for much smaller labels

| # of pts: | Label dimensions (in pixels) | | |
|---|---|---|---|
| | 16x4 | 3x1 | 1.0x0.4 |
| 11K | 0.110 | 0.110 | 0.093 |

These results compare favorably against all previously published methodologies. In terms of speed, this algorithm appears to be orders of magnitude faster than most other approaches, including those documented in [9, 20, 27, 32, 36]. Furthermore, these results include tests ranging up to 130,000 features, whereas no published results to date include tests with datasets larger than 20,000 features, with most less than 5000. (The implementation in [2] mentions a dataset of 12 million, but no timed results are offered).

In order to compare this algorithm with previously reported approaches, it was also applied to some of the benchmark data available at [33]. It is important to note, however, that direct comparisons are difficult due to a fundamental design difference, viz: nearly all previous approaches assume a fixed-size map resolution and allow no features to be obscured. In contrast, this paper has described an algorithm suitable for maps of varying-resolutions, in which point-density *necessitates* feature overposting on most zoom levels. Moreover, previous approaches typically measured the effectiveness of an algorithm by the number of features labeled in the final solution. This metric is clearly unsuitable, however, when applied to a map that is so dense that the vast majority of its points cannot possibly be labeled. Nevertheless, some comparisons are indeed useful. The fastest recorded times previously reported in the literature appear in [27]. Although the authors of that approach do not specify the speed of the system it was run on, these results, tabulated in **Table 6,** indicate that the trellis strategy may be up to ten times faster. Note that this is true even though the trellis algorithm was processing data sets that were sometimes ten times larger.

**Table 6:** Benchmark Data

| | US Cities | | |
|---|---|---|---|
| | No. of sites | # of labels | time (in sec) |
| Algorithm in [27]. | 1,041 | 1,041 | 0.5 |
| Trellis algorithm | 10,296<br>100,000<br>130,000 | 1,537<br>6,883<br>63,269 | 0.078<br>0.86<br>1.3 |
| | **Munich Drill Holes** | | |
| Algorithm in [27]. | 19,461 | 11,049 | 7.1452 |
| Trellis algorithm | 19,446 | 11,142 | 0.204 |

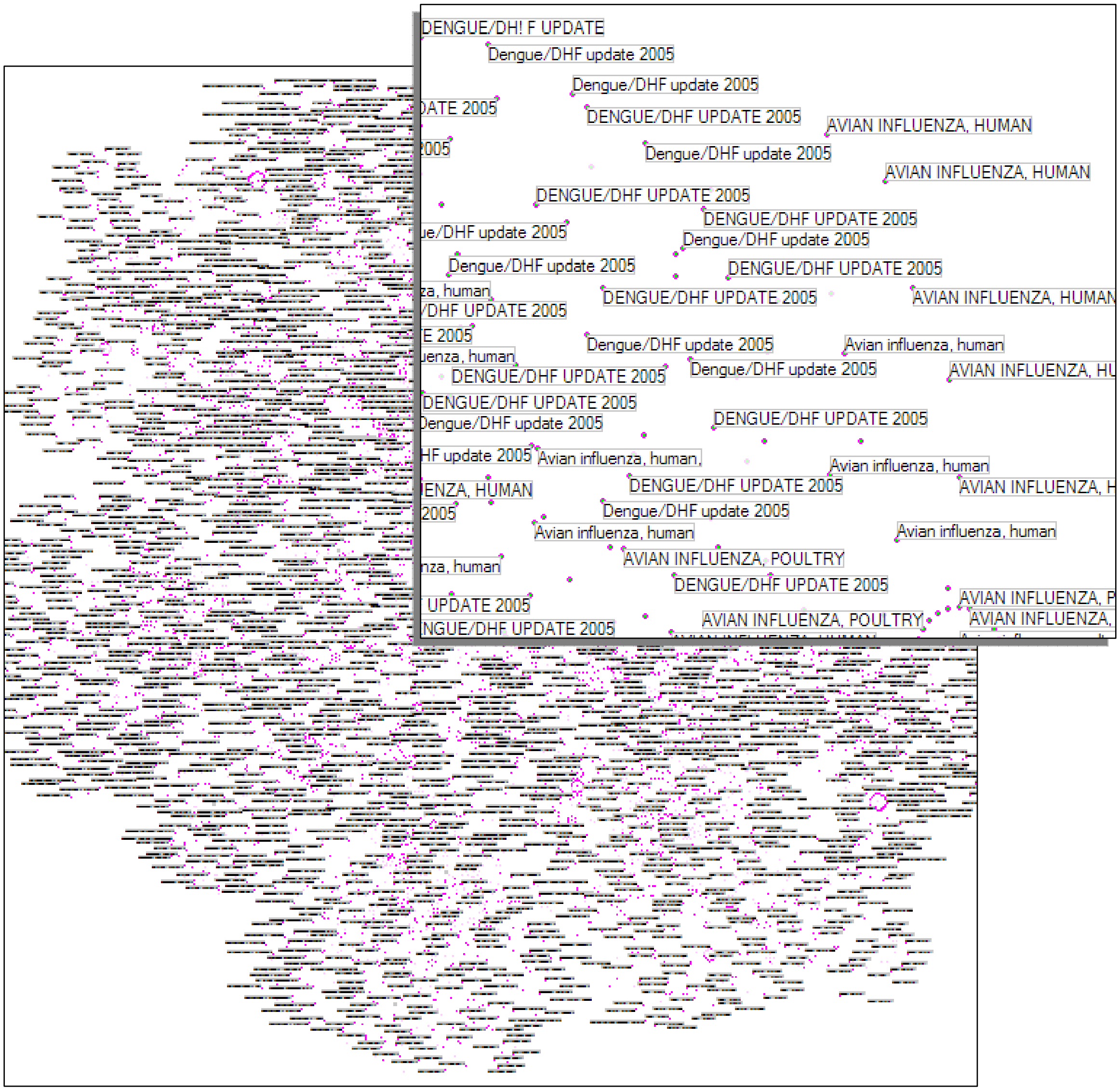

**Figure 4:** 11,000 feature points from a cluster graph. At the given resolution 3,605 labels were placed in 0.11 seconds. This paricular view is not meant for display, but demonstrates the caluclated label positions of one "zoom level." Eight other such zoom levels can be calculated in under a second, allowing the user to zoom in until all labels are deconflicted. The inset shows what the user would see at the current zoom level.

## 6 Conclusion and Suggestions for Further Study

This paper has presented a new approach for automated feature label de-confliction operating at speeds and scalability that are well-suited for interactive visual analytic applications. An algorithm was offered that begins to realize Dr. Freeman's dream of a "1-second quality labeling of an electronically displayed map" [12]. Indeed, the approach outlined here provides real-time, whole-map labeling at speeds measured in milliseconds, without the need for preprocessing. Moreover, this method has demonstrated an ability to scale, in sub-second time, to massive data sets, larger than what most previous approaches have even attempted to handle. This is a critical feature at a time when visual analytic applications routinely process tens of thousands of nodes in a single view.

The speed and scalability of this algorithm opens the door to a number of options in terms of 2D and 3D interactivity. For the two-dimensional case, two distinct modes of operation are possible:

*"Just-in-time" view de-confliction.* The original motivation for this study was to produce a de-confliction algorithm fast enough to operate at a rate of multiple frames per second. This goal has indeed been realized with data sets in excess of 25,000 nodes. At these speeds a view can be efficiently labeled and re-labeled at every interactive movement of a user's mouse. This was, in fact, the original design of the algorithm: for any given orientation of the data, the *current view* can be labeled and any previous labeling can be discarded. One of the drawbacks of this approach, however, is its failure to satisfy the so-called "desiderata," or rules of label consistency (outlined in [2]), which seek to eliminate unexpected "popping" and re-sorting of label locations. For this reason, a second option can be considered:

*Multi-level pre-processing.* This paper has, from the outset, presented pre-processing as inappropriate for the ad hoc nature of visual analytic maps. Nevertheless, the speed of this algorithm allows us to reconsider that opinion. Recall that camera zooming can be considered equivalent to a universal expansion of intrapoint distances along with label-size scaling. Therefore, by pre-scaling the labels to progressively smaller sizes, the layout configuration may be computed for every zoom level at construction time. As seen in Table 5, such computations add only minimal expense. Hence, in 3-5 seconds, an entire map could be pre-processed, and its label locations stored, for up to 8 or 12 zoom levels. In so doing, the undesirable "popping" of labels that would otherwise occur as the user pans around in lower zoom levels would be eliminated. The added memory demands can be addressed through subdivision of the trellis, and the small amount of extra time required would in many cases be dwarfed by the demands of view construction in general. This would also free up cycles that may be better spent by the expensive rendering engine during view interaction. It should be said that, while this would prevent any "popping" during horizontal movement (panning), other measures are necessary to mitigate popping between zoom levels. One option in this regard would be to "lock" label locations in place, once they have been determined at a higher zoom level. Such a decision would expose the unavoidable trade-off between label consistency and label-count maximization.

Beyond the advantages offered for labeling in the two-dimensional case, this algorithm may also be applicable in the more demanding arena of three-dimensional views. Heretofore, no labeling algorithm has offered the speed required to handle the complexities of interactive 3D label de-confliction. When a user interacts with a three-dimensional view, the relative orientation and configuration of the features in the view are constantly in flux: the 2D zooming and panning capabilities are now supplemented with rotation (both camera and view), view angle manipulation, and dollying. The possible number of view orientations is literally countless and, hence, a pre-processing phase is unfeasible. Yet, by projecting the features to the view-plane, the multi-frame-per-second rate of this algorithm may finally provide real-time labels for three-dimensional views.

Another area for future study presents itself: Because the trellis strategy operates only on specific regions within a two-dimensional array, it readily lends itself to a threaded, parallel-processing approach. In fact, the algorithm may well be classified as an "embarrassingly parallel" problem—easily separable into independent tasks. See [13], particularly chapter 8, for a description of how this might be accomplished. But even without that, it is clear, that for a vast number of applications within the arena of information visualization, the trellis strategy of label de-confliction is a fast, reliable, and worthwhile tool.

#### ACKNOWLEDGEMENTS

The author wishes to thank many colleagues for their early reviews and valuable comments, including Pak Chung Wong, Jean Scholtz, Lee Ann Dudney, and John Miller, as well as the many helpful comments of the anonymous reviewers. Particular thanks are due to John Risch whose ideas and encouragement provided inspiration to this project.